\documentclass[12pt]
{article}
\usepackage{amsfonts,amssymb,amsmath,amsthm} 
\usepackage{epigraph}
\title{
Effective algorithms for homogeneous utility functions.}
\author{Alexander Shananin
\thanks{{\tt e-mail: alexshan@yandex.ru} Supported in part by RFBR grant ¹ 17-01-00507
}
\and
Sergey Tarasov
\thanks{{\tt e-mail: serge99meister@gmail.com} Supported in part by RFBR grant ¹ 17-07-00300
}
}

\date{ }
\newtheorem{theorem}{Theorem}

\newtheorem{rK}{Remark}

\newtheorem{proposition}{Proposition}
\newtheorem {corollary}{Corollary}

\let\eps\varepsilon
\def\RR{{\bf R}}

\def\F{{\cal F}}

\def\P{{\cal P}}

\def\BB{{\bf B}}

\begin{document}

\maketitle

\epigraph{ tweets: I am an economist so I can ignore computational constraints.

I am a computer scientist, so I can ignore gravity.
}{ Lance Fortnow, a computer scientist}

\begin{abstract}
Under the assumption of (positive) homogeinity (PH in the sequel)
of the corresponding utility functions, 
we construct polynomial time algorithms for  the weak separability, 
 the collective consumption behavior 
 and some related problems. These problems are known to be at least $NP$-hard if the homogeinity assumption is dropped.

\end{abstract}

\let\es\varnothing
\let\ph\varphi
\let\eps\varepsilon
\let\al\alpha
\def\Tb{T^\text{\textup{bad}}}
\def\Ta#1{T^{\text{\textup{all},\:}#1}}
\def\P{\cal P}
\def\F{\cal F}
\def\C{\cal C}
\def\HC{\cal CH}
\def\M{\cal M}
\def\BB{\mathbb B}
\def\CC{\mathbb C}
\def\RR{\mathbb R}
\def\SS{\mathbb S}
\def\ZZ{\mathbb Z}
\def\NN{\mathbb N} 
\def\FF{\mathbb F}
\def\QQ{\mathbb Q}
\def\fI{\ensuremath{I_\BB}}
\def\fP{\ensuremath{P_\BB}}
\def\fS{\ensuremath{S_\BB}}
\def\id{\mathrm{id}}
\def\reg(#1){$#1$-reali\-za\-bi\-li\-ty}
\def\Per{\ensuremath{\mathrm{Per}}}
\def\PePe{(\Per_{\Sigma_1}{\parallel}P_{\Sigma_2})}
\def\pepe{\Per_{\Sigma_1}{\parallel}P_{\Sigma_2}}

 \section{Introduction}

 Globalization of the world economics results in structural changes that seems to fell out of the scope of the traditional indicators such as, for instance, domestic product, and it seems that new instruments like statistics panel should be involved in analysis. In this setting some already studied problems of economics theory are becoming highly relevant. This paper is mainly motivated by the two following questions.
 
\smallskip

1. Is it possible to describe structural changes within the framework of the traditional Pareto's theory of rational representative consumer (or its generalizations) for the case of several consumers?

2. How to analyze the segmentation of consumer's market and how to compute statistics panel? In economics literature this question is often referred to as separability problem.

\smallskip

Recall that Pareto's theory assumes that the demand can be described within the model of a rational representative consumer maximizing his utility function under the budget constraints  wherein the utility is postulated to be {\em well-behaved}, i.~e. it is assumed to be {\em concave, increasing,  non-satiated 
(nonnegative)} maps from $\RR^n_+$ to $\RR_+$ {\em continuous up to the boundary}.  Within the framework of economic indices  it is necessary to make another requirement and postulate that utility is PH. Indeed (see, e. g. \cite{Kond}), money balance of the consumer goods in the initial range and in the aggregated indicators is equivalent to Euler's identity for the utility function. On the other hand,  in economic literature PH-assumption is criticized both from theoretical and empirical grounds. In theory, the Engel curves for the PH-utility function should be rays, and this contradicts the intuition on demand saturation with the growth of income of the household. Empirical studies of individual households also show that their Engel curves are not rays.

Nevertheless, in some important settings PH can be justified. For instance,  in the neoclassical theory of consumer demand PH-assumption  is verified in \cite{Shananin}. Namely, it is reasonable to assume that PH holds not for an individual household but for the whole group of households with comparable incomes (in other words, groups with similar living standards). Especially, the size of a particular group varies as changes of income of an individual household result in modification of its consumer behavior and possible transitions to different groups. This pattern leads to an essential nonlinearity of Engel curves. And it turns out that the validity of PH-assumption for the cumulative consumer behavior strongly depends on the redistribution of prices under changes of incomes of social groups, and, in particular, whether these changes could be described by the Bergsonian social welfare function. 

In this paper for the case of PH-utilities 
we construct effective algorithms bypassing exaustive search for some important problems of the revealed preference theory that are NP-hard in the nonhomogeneous setting. In this respect the main message of the paper is that  models allowing for PH utility can be analyzed essentially faster, which may be essential for, say, monitoring or similar purposes.

The organization of the paper is as folows. In the next section we recall some standard facts of the revealed preference theory. The third and the fourth sections are devoted to the desciption and justification of the algorithms for the weak separability and, respectively, for the  collective consumption behavior problems.

 For convenience the details of some proofs are moved to the Appendix.

\section{The revealed preference theory}

\hspace{4mm} Let us recall some standard facts from the revealed preference theory.

The rationality problem consists in the following: {\bf given} the market statistics for a certain period of time
$S=\{\{p^t,q^t\},\, p^t=(p^t_1,\dots,p^t_n)\in \RR^n_+ ,\, q^t=(q^t_1,\dots,q^t_n)\in \RR^n_+,\, t=1,\dots,T\}$ (here $p^t$ and $q^t$ are the bundles of prices and the corresponding quantities of goods purchased at time $t$, respectively), {\bf to find}
whether the statistics can be {\em rationalized}, where 
rationalization means the existence of a certain {\em (concave) utility function} $f: \RR^n_+\to\RR_+$
usually called {\em the index of goods}, which is {\em consistent} with $S$.
By definition, a utility function $f(\cdot)$ is consistent with market statistics $S$ if
$$q^t \in \mbox{Argmax }\{f(q)\,\mid\, p^t q \leq p^t q^t, \, q \geq 0\},
\quad t=1,\dots,T.$$

Hereinafter we assume that utility functions are {\em well-behaved}, i. e. they are assumed to be {\em concave, increasing,  non-satiated 
(nonnegative)} maps from $\RR^n_+$ to $\RR_+$ {\em continuous up to the boundary}.  Besides that we assume that utilities satisfy an additional more restrictive 
reqiurement of {\em positive homogeneity (PH)}\footnote{This property is also referred to as {\em homotheticity} in economics literature, but it seems that our choice is more suitable, as not only level sets of the PH utilities ought to be homothetic, but, e. g., Euler identity should also hold.}.  The results of this paper may provide a convincing argument in favour of this choice, if the model involved allows for it: if we drop homogeneity assumption then all the problems listed below are hard from the point of the complexity theory as they are known to be at least $NP$-hard and may be even harder, say, belong to the complexity class $\exists\RR$-hard. 
Additionally, numerical results (see, for instance, \cite{ShanKlem}) show  that the usage of PH utility functions are more adequate and produce essentialy less noise. 

Recall that by the {\bf homogeneous Afriat's theorem} (see, e.g. \cite{Afriat, Varian}) 
the statistics $S$ is rationalizable in the class of the  well-behaved PH utilities iff the following
system of linear inequalities is consistent
\begin{equation}\label{system}
\lambda_t\, p^t q^t \leq \lambda_\tau
\, p^{\tau} q^t, \, \lambda_t > 0, \, \lambda_\tau > 0,\, t,\tau = 1, \ldots, T.
\end{equation}
Evidently, we can assume that the additional normalization condition holds $\sum_{i=1}^T \lambda_i=1$.

The particular utility function can be
recovered from an
arbitrary solution of (\ref{system}) by the formula
\begin{equation}\label{utility}
f(x) =\min_{t=1, \ldots,T} \lambda_t \, p^t x.
\end{equation}

Now we remind two fundamental problems of the revealed preference theory and prove 
that for PH ulilities each of these problems can be solved effectively by a polynomial time algorithm. 
On the other hand, under the standard setting not assuming PH, 
these problems are proven to be at least $NP$-hard, i.~e. intractable for large enough inputs.

\section{Weak separability}

Assume that goods are partitioned into two (non intersecting) groups of, say, $k$ q-goods and $l$ y-goods.
Then the
market statistics bundle can be written as $S=\{p^t, x^t; q^t, y^t\}_{t=1,\dots, T}$, here $p^t, x^t$ are prices and $q^t, y^t $ are the corresponding volumes of purchases and thus
$\{p^t; q^t\}_{t=1,\dots, T}$ is a market statistics bundle for q-goods and, 
respectively, $\{x^t; y^t\}_{t=1,\dots, T}$ is a market statistics bundle 
for y-goods. 

Assume that the overall statistics $S$  is rationalizable by a well-behaved utility function $u({\mathbf q,\,{\mathbf y}})$, where ${\mathbf q}=\{ q^t\},\, {\mathbf y}=\{ y^t\},\, t=1,\dots, T$.

We say that $S$ is rationalizable by weak separability (in y-goods) if there exist a well-behaved ``micro'' subutility function $v({\mathbf y})$ rationalizing statistics for y-goods  and  a  well-behaved ``macro (aggregated)'' utility function $\tilde u({\mathbf q}, v(y))$, strictly increasing in $v$-variable, such that $u({\mathbf q,\,y})=\tilde u({\mathbf q}, v(y))$.

This problem is studied in many publications, see, for instance, \cite{BPR-separab, Varian, SW-1994, FW-2008, ch2015}.

Necessary and sufficient conditions for  (non homogeneous) weak separability are proven, e. g., in \cite[Theorem 3]{Varian}. Furthemore,  \cite[Theorem 5]{Varian} gives necessary and sufficient conditions for the case of the so called ``homothetic separability'', where the subutility function $v(y)$ is additionally assumed to be PH (homothetic). Unfortunately, both characterizations are expressed by a set of quadratic inequalities and  moreover are intractable from the point of view of complexity theory. It is proved in \cite{ch2015, Echen-2014} that testing weak separability is $NP$-complete. 

We prove that if we assume {\bf complete PH-separability}, i. e. that not only subutility is PH but macro utility is also PH, then testing weak separability can be reduced to convex program that can be solved by
an effective polynomial time algorithm. 

The following necessary and sufficient conditions for complete PH-separability can be obtained along the same lines as the corresponding conditions for
 homothetic separability in \cite[Theorem 5]{Varian}. 
 For convenience the details of the proof are moved to the Appendix.

\begin{eqnarray}
\lambda_t\, x^t y^t \leq \lambda_\tau
\, x^{\tau} 
y^t,\,&& \lambda_t > 0, \, \lambda_\tau > 0\label{f1}\\
\mu_{t} \lambda_\tau (\sum_{i=1}^{k}  p^t_i \, q^t_i+  \sum_{i=1}^{l}  x^t_i \, y^t_i)&\leq &
\mu_{\tau} (\lambda_\tau \sum_{i=1}^{k}  p^\tau_i \, q^t_i+  \lambda_{t}  \sum_{i=1}^{l}  x^t_i \, y^t_i) \label{f2}\\
\sum_{i=1}^T \lambda_i&=&1\label{f3}\\
t,\tau &=&1,\dots, T. 
\label{f4}
\end{eqnarray}

Let us show  how to put the system involved into  convex form.

For convenience we use capital letters for all constant parameters except indices and denote the logarithm of some [positive] parameter $x$ by $\tilde x$. For instance, $\widetilde{{}^\alpha\lambda_t}\stackrel{def}=\log {}^\alpha\lambda_t$. We rewrite the system using this notation.

\begin{equation*}
\tilde\lambda_t + \log(X^t Y^t) \leq \tilde\lambda_\tau +
\log(X^{\tau} 
Y^t) 
\end{equation*}
\begin{align}
\tilde\mu_{t} + \tilde\lambda_{\tau} + \log (\sum_{i=1}^{k}  P^t_i \, Q^t_i+  \sum_{i=1}^{l}  X^t_i \, Y^t_i)\leq 
\tilde\mu_{\tau} + \\ + \log [\exp(\tilde \lambda_\tau) \sum_{i=1}^{k}  P^\tau_i \, Q^t_i+
\exp(\tilde \lambda_{t}) \sum_{i=1}^{l}  X^t_i \, Y^t_i] \label{g2}
\end{align}
\begin{eqnarray}
\sum_{i=1}^T \exp(\tilde \lambda_i)&=&1\label{g3}\\
t,\tau &=&1,\dots,T. 
\nonumber 
\end{eqnarray}

Now the lhs of \eqref{g2} is already convex. To fix the rhs  we substitute for each  $\lambda_i$ its expression from the normalization condition \eqref{f3}. We obtain a function of the form $\log[const-\sum_r\exp (L_r)]$, where $L_r$ are affine forms of $\widetilde{\lambda_i}$ and thus the function involved is concave. At last we need to fix the problem with the normalization equality \eqref{f3} and this is done by adding  a new nonnegative slack variable $\gamma$. 
Finally, we get the convex program

\begin{eqnarray}
\gamma &\to & \min \\
\tilde\lambda_t + \log(X^t Y^t) &\leq& \tilde\lambda_\tau +
\log(X^{\tau} 
Y^t)
\end{eqnarray}
\begin{align}
\tilde\mu_{\tau} + \tilde\lambda_t + \log (\sum_{i=1}^{k}  P^t_i \, Q^t_i+  \sum_{i=1}^{l}  X^t_i \, Y^t_i)\leq 
\tilde\mu_{\tau} + \log [(1-\sum_{i=1,\,i\neq\tau}^T\exp(\tilde \lambda_i)) \sum_{i=1}^{k}  P^\tau_i \, Q^t_i+ \\ 
(1-\sum_{i=1,\,i\neq t}^T\exp(\tilde \lambda_{i})) \sum_{i=1}^{l}  X^t_i \, Y^t_i] 
\end{align}
\begin{eqnarray}
\sum_{i=1}^T \exp(\tilde \lambda_i)+\gamma&\leq&1, \,\gamma\geq 0 
\\
t,\tau&=&1,\dots,T. \nonumber 
\end{eqnarray}

And we conclude that the statistics $S$ is weakly separable iff the optimum of the related convex program  is zero. Moreover, we can use modern effective polynomial-time routines to solve this program. To be a bit more exact by ``effective'' 
we mean a procedure which solves the approximate version of the convex 
program involved within the accuracy $\eps$ in time $poly(\max(L,\log\frac{R}{\eps}))$, where $L$ is  the length of the input and $R$ is the so called localization radius. By definition, if the program is consistent then at least one solution should fall into the ball of radius $R$ with the center at the origin.

By construction we have proved the following theorem.

\begin{theorem}
There is an effective polynomial time algorithm to solve 
weak separability problem 
\end{theorem}

\begin{rK}
PH-separability is equivalent to the so called {\bf indirect weak separability (IWS)} (see, e.~g. \cite{BS-2000, ch2015}) and as a by product IWS for PH utilities can be solved by an effective algorithm.
\end{rK}

Let us define the {\bf generalized separability problem} as a problem of  checking whether there {\em exists} some separable partition of the market statistics\footnote{Evidently, to avoid trivial solutions we should assume that the required partition should be balanced, i. e., comparable to the size of the statistics.}. 

\begin{corollary}
The generalized PH-separability problem belongs to $NP$.
\end{corollary}

\section{Collective rationality}

Afriat's theorem mentioned above is an effective tool to test whether the market 
statistics can be rationalized. But how could violation of rationality be dealt, 
for instance, in the case of some essential ongoing structural changes?
In  view of the uprecedented and unexpected changes of the modern economics this 
question seems to acquire the status of a real challenge of crucial importance. 
One of the approaches to deal with this problem is the collective revealed 
preference rationality model introduced in \cite{ch2007}. 
It consists in partitioning of the initial (nonrationalizable) market statistics 
bundle into a certain number of, say $k$, rationalizable bundles 
of the same cardinality, having the same initial price subbundles but 
different goods subbundles that componentwise sum to the initial goods subbundle. 
In other words, we want to represent the utility as a $k$-dimensional vector 
function  such that each
component 
is a well-behaved PH utility function for the statistics 
$S_\alpha\stackrel{def}=\{p^t,{}^\alpha q^t\},\, t=1,\dots,T$, where for all 
$i\in \{1,\dots,n\}$ $\sum_{\alpha=1}^k {}^\alpha q^t_i=q^t_i$. Using analogy with smooth setting\footnote{In  \cite{Shananin} this parameter is related to the E. Cartan's class number of the differential form of the demand. Loosely speaking, it is equal to the minimal number of the independent variables (functions) needed to express the form.} we call minimal such $k$ the {\em discrete class number of the demand form}.

By the homogeneous Afriat's theorem\footnote{This problem becomes intractable 
\cite{fab2010} if we drop homogeneity assumption (and use the standard Afriat's 
conditions).} we can express collective rationality by the following system. 
\begin{eqnarray*}
{}^\alpha\lambda_t \sum_{i=1}^n  p^t_i \,\,{}^\alpha q^t_i&\leq &
{}^\alpha\lambda_\tau \sum_{i=1}^n  p^\tau_i \,\,{}^\alpha q^t_i, \quad \alpha=1,\dots k\\
\sum_{\alpha=1}^k {}^\alpha q^t_j&=&q^t_j,\quad j=1,\dots,n,\\
t,\tau&=&1,\dots,T.
\end{eqnarray*}

As above, we use capital letters for all constant parameters except indices and denote the logarithm of some [positive] parameter $x$ by $\tilde x$.
After taking logarithms we obtain
\begin{eqnarray}
\widetilde{{}^\alpha\lambda_t} +\log[\sum_{i=1}^n P^t_i\exp(\widetilde{{}^\alpha q^t_i})]&\leq& \widetilde{{}^\alpha\lambda_\tau}+
\log[\sum_{i=1}^n  P^\tau_i\, {}^\alpha q^t_i],\quad \alpha=1,\dots k\label{main1}\\
\sum_{\alpha=1}^k {}^\alpha q^t_j&=&Q^t_j,\quad j=1,\dots,n,\label{main2}\\
t,\tau&=&1,\dots,T.
\nonumber
\end{eqnarray}

Now we can use almost the same trick as before to transform the resulting system  into a convex program. 

Note that the lhs of \eqref{main1} is already convex. To fix the rhs  we substitute for each  ${}^\alpha q^t_i$ its expression from the relevant balance condition from \eqref{main2}. We obtain a function of the form $\log[const-\sum_r\exp (L_r)]$, where $L_r$ are affine forms of $\widetilde{{}^\alpha q^t_i}$ and thus the function involved is concave. At last we need to fix problem with balance equalities \eqref{main2} and this is done by adding  new nonnegative slack variables $\gamma^t_i,\, t=1,\dots,T,\, i=1,\dots,n$. Finally, we get the convex program

\begin{eqnarray*}
\sum_{t=1}^T\sum_{i=1}^n \gamma^t_i &\to& \min\\
\widetilde{{}^\alpha\lambda_t} +\log[\sum_{i=1}^n \exp(\widetilde{P^t_i}) \exp( \widetilde{{}^\alpha q^t_i})] &\leq & \widetilde{{}^\alpha\lambda_\tau}+
\log[\sum_{i=1}^n  P^\tau_i (Q^t_i-\sum_{j=1,\,j\neq i}^n \exp(\widetilde{{}^\alpha q^t_j}))]\\
\sum_{\alpha=1}^k \exp(\widetilde{{}^\alpha q^t_i})+\gamma^t_i &\leq& Q^t_i, \quad \alpha=1,\dots k,\\ 
\gamma^t_i\geq 0\\
t,\tau=1,\dots,T,& & i=1,\dots,n.
\end{eqnarray*}

By construction we proved the following theorem.

\begin{theorem}
The discrete class number of the demand form is less or equal to $k$ iff the resulting convex program has optimal solution zero.
\end{theorem}

As above we can use modern effective polynomial-time routines to solve the convex problem involved for any $k$ and
thus   not only can we identify the minimal $k$ but we are able to retrieve some 
required utility vector function as well.

\section{Appendix}

\begin{proposition}
Formulas \eqref{f1}--\eqref{f4}  give necessary and sufficient conditions for
complete PH-separability.
\end{proposition}

{\em Proof.} At first, recall that Euler's homogeneous function theorem can be reformulated as follows for the case of PH well-behaved functions $f(\cdot): \RR^n_+\to\RR_+$. If $x_0\in \RR^n_+$ and $p\in \partial f(x_0)$ then $p x_0 = f(x_0)$.

{\em Necessity.} In the notation used above we need to prove the existence of the well-behaved PH functions $u_0(q,z)$ and $u_1(y)$ such that 

\begin{equation}\label{crit}
(q^t,y^t)\in Argmax[u_0(q, u_1(y))\,\mid\, p^t q + x^t y\leq p^t q^t + x^t y^t,\, q\geq 0, y\geq 0],\, t=1,\dots,T.
\end{equation}

As PH well-behaved functions are monotonic then it follows from \eqref{crit} that
$
y^t\in Argmax[u_1(y)\,\mid\, x^t y\leq x^t y^t,\, y\geq 0],\, t=1'\dots,T.
$

Define the Young transforms: $\nu_0(p,s)\stackrel{def}{=} \inf\limits_{\{q\geq 0,\,z\geq 0\,\mid,\, u_0(q,z)>0\}} \frac{p q + s z}{u_0(q,z)}$ and  $\nu_1(x)\stackrel{def}{=} \inf\limits_{\{y\geq 0\,\mid\, u_1(x)>0\}} \frac{x y}{u_1(x)}$. By definition it holds
\begin{eqnarray*}
\nu_1(x^t) u_1(y^t) &=& x^t y^t\\
\nu_1(x^{\tau}) u_1(y^t) &\leq& x^{\tau} y^t\\
\nu_0(p^t,\nu_1(x^t))u_0(q^t, u_1(y^t)) &=& p^t q^t + \nu_1(x^t) u_1(y^t)\\
\nu_0(p^\tau,\nu_1(x^\tau))u_0(q^t, u_1(y^t)) &\leq& p^\tau q^t + \nu_1(x^\tau) u_1(y^t)\\
t, \tau &=& 1,\dots, T.
\end{eqnarray*}

Now necessity follows as $\lambda_t=\frac1{\nu_1(x^t)}, \, \mu_t=\frac1{\nu_0(p^t, \nu_1(x^t))}\,\, t=1,\dots, T$ are solutions of the system \eqref{f1}--\eqref{f4}.

{\em Sufficiency.} Set $u_1(y)\stackrel{def}{=} \min\{\lambda_t x^t\,\mid\, t=1,\dots, T\}$. By definition $y^t\in Argmax[u_1(y)\,\mid\, x^t y\leq x^t y^t\,, y\geq 0], \, t=1,\dots, T.$ Set $u_0(q,z)\stackrel{def}{=} \min\{\mu_t\left(p^t q + \frac1{\lambda_t}z\right) \,\mid\, t=1,\dots, T\}$.

Let $q\geq 0,\, y\geq 0,\, p^t q + x^t y\leq p^t q^t + x^t y^t$. Then it holds

\begin{eqnarray*}
p^t q^t + x^t y^t = p^t q^t + \frac1{\lambda_t}u_1(y^t) &=&  \frac1{\mu_t}u_0(q^t, u_1(y^t)),\\
x^t y\geq \frac1{\lambda_t}u_1(y),\, p^t q + x^t y\geq p^t q +\frac1{\lambda_t}u_1(y) &\geq& \frac1{\mu_t}u_0(q, u_1(y)).
\end{eqnarray*}

And it follows that $u_0(q^t, u_1(y^t))\geq u_0(q, u_1(y))$ and thus \eqref{crit} holds.


\begin{thebibliography}{99}

\bibitem{Afriat}
Afriat S. The
construction of utility functions from expenditure data. //{\em
International Economic Review}.  8. N1. 66--67. 1967.

\bibitem{BPR-separab} Blackorby C., Primont D., and Russel R.
Duality, Separability, and Functional Structure: Theory and Economic Application.
Amsterdam: North-Holland. 1979.



\bibitem{BS-2000} Brown D.J., Shannon  C. Uniqueness, stability, and comparative statics in
rationalizable Walrasian markets. Econometrica 68, 1529--1540. 2000.

\bibitem{ch2007}  Cherchye L.,  De Rock B., and Vermeulen F. The Collective model of household consumption: a non parametric characterization. Econometrica. Vol. 75. No. 2. 553–-574. 2007.

\bibitem{ch2015} Cherchye, L., Demuynk, T., De Rock, B., Hjertstrand P. Revealed preference tests  for weak separability: an integer programming approach. Journal of Econometrics. 186. 129--141. 2015.


\bibitem{Echen-2014} Echenique F. Testing for separability is hard.  	arXiv:1401.4499 [cs.GT]

\bibitem{HH-2017} Heufer J., Hjertstrand P. Homothetic Efficiency: Theory and
Applications, Journal of Business \& Economic Statistic. DOI:
10.1080/07350015.2017.1319372. 2017


\bibitem{FW-2008} Fleissig A., Whitney G.A. A nonparametric test of weak separability and
consumer preferences. Journal of Econometrics. 147, 275–281. 2008.

\bibitem{ShanKlem} Klemashev N., Shananin A. 
Inverse problems of demand analysis and their applications to computation of positively-homogeneous Kon$\ddot{ \mbox{u}}$s–Divisia indices and forecasting. Journal of Inverse and Ill-Posed Problems. 2015.

\bibitem{Kond} Kondrakov I., Pospelova L., and Shananin A. Generalized nonparametric method. Applications to the analysis of commodity markets. In Russian. Proc. MIPT 2. 32–45. 2010.

\bibitem{Shananin} Petrov A., Shananin  A. Integrability conditions, Income distribution, and social structures. Lecture notes in Economics and mathematical systems, 453. Springer. 1998.

\bibitem{shat2015} Shananin A., Tarasov S. Computing the class the differential demand form from discrete data. Proc. MIPT conference. 2015.

\bibitem{sme2015} Smeulders, B., Cherchye, L., De Rock, B., Spieksma, F., Talla Nobibon, F. Complexity results for the weak axiom of revealed preference for collective consumption models. Journal of Mathematical Economics. 58. 82--91. 2015.

\bibitem{SW-1994} Swofford J.L., Whitney, G.A. A revealed preference test for weakly separable
utility maximization with incomplete adjustment. Journal of Econometrics. 60, 235–249.  1994.


\bibitem{fab2016} Talla Nobibon, F., Cherchye L., Crama, F,  Demuynk, T., De Rock, B., and Spieksma, F.
Revealed preference tests of collectively rational consumption behavior: formulations and algorithms. Operations Research 64, 1197–-1216. 2016.




\bibitem{fab2010} Talla Nobibon, F., Spieksma, F. On the complexity of testing the collective axiom of revealed preference. Mathematical Social Sciences. 60. No. 2. 123--136. 2010.





\bibitem{Varian}
Varian H.R.  Non-parametric tests of consumer
behaviour. //{\em The Review of Economic Studies}. 1983, 50(1).
99--110. 1983.






\end{thebibliography}
\end{document}